%% file: main.tex
\documentclass[%
reprint,
superscriptaddress,
nofootinbib,
amsmath,amssymb,
aps,
pre,
notitlepage,
showkeys]{revtex4-2}
\usepackage[colorlinks=true, urlcolor=blue, anchorcolor=blue, citecolor=blue,filecolor=blue,linkcolor=blue,menucolor=blue]{hyperref}
\usepackage{url}
\usepackage{graphicx}
\usepackage{lipsum}
\usepackage[export]{adjustbox}
\usepackage{mathtools}
\usepackage{gensymb}
\usepackage[T1]{fontenc}
\usepackage{amsmath}
\usepackage{amssymb}
\usepackage{stmaryrd}
\usepackage{fancyhdr}
\usepackage[utf8]{inputenc}
\usepackage{rotating}
\usepackage{comment}
\usepackage{url}
\usepackage{bbm}
\usepackage{soul}
\usepackage[normalem]{ulem}
\usepackage{tabularx,booktabs}
\usepackage{stmaryrd}
\newcolumntype{Y}{>{\centering\arraybackslash}X}

\usepackage{hyperref}  %hyperref still needs to be put at the end!
\usepackage{subfigure}

\begin{document}
\title{First-passage on disordered intervals} 
\author{James Holehouse}
\email{jamesholehouse@santafe.edu}
\affiliation{The Santa Fe Institute, 1399 Hyde Park Road, Santa Fe, NM, 87510, USA}

\author{S.~Redner}
\email{redner@santafe.edu}
\affiliation{The Santa Fe Institute, 1399 Hyde Park Road, Santa Fe, NM, 87510, USA}

\begin{abstract}
We investigate the first-passage properties of nearest-neighbor hopping on a finite interval with disordered hopping rates. We develop an approach that relies on the backward equation, in conjunction with probability generating functions, to obtain all moments, as well as the distribution of first-passage times.  Our approach is simpler than previous approaches that are based on either the forward equation or recursive method, in which the $m^{\rm th}$ moment requires all preceding moments.  For the interval with two absorbing boundaries, we elucidate the disparity in the first-passage times between different realizations of the hopping rates and also unexpectedly find that the distribution of first-passage times can be \emph{bimodal} for certain realizations of the hopping rates.
\end{abstract}

%\date{\today}
\maketitle

% \section{Introduction}
First-passage problems concern the distribution of times for a stochastic process to first reach a defined state~\cite{feller1991introduction,van1992stochastic,karlin2014first,redner2001guide,bray2013persistence}. A paradigmatic setting for first-passage phenomena is the nearest-neighbor random walk on the finite interval. Here, it is of fundamental importance to understand the statistics of the time required for the random walk to \emph{first} reach the boundaries of the interval. The applications of this type of first-passage problem are vast, including biological processes, such as the Moran model~\cite{moran1958random,antal2006fixation}, migration phenomena~\cite{mckane2000mean,pinero2022fixation}, the behavioral dynamics of ant recruitment~\cite{kirman1993ants,holehouse2022exact}, as well as the dynamics of many types of financial instruments~\cite{perello2011scaling,chicheportiche2014some}.

First-passage phenomena are much richer and less well understood when the hopping rates of the random walk are \emph{spatially disordered}. While much progress has been made in determining the average first-passage time and its low-order moments~\cite{van1992stochastic,redner2001guide,ashcroft2016metastable,PhysRevA.40.2082,PhysRevA.42.4503,PhysRevA.41.4562,PhysRevE.49.R967,goldhirsch1986analytic,noskowicz1990first}, understanding the properties of the full \emph{distribution} of first-passage times is still incomplete. Here we tackle this problem by focusing on the generating function for the first-passage probability. The finite interval provides a particularly instructive platform to investigate many basic disorder-controlled physical phenomena, such as the diffusion of a particle in a random potential~\cite{sinai1983limiting,goldhirsch1986analytic,noskowicz1990first,bouchaud1987relaxation,fisher1998random,le1999random,krapivsky2010kinetic}, DNA translocation through a nanopore~\cite{slutsky2004diffusion}, channel transport~\cite{iyer2016first}, and percolation \cite{lyons1992random,mardoukhi2018fluctuations}. Our formalism allows us to readily evaluate first-passage probabilities and their moments for individual realizations of disorder and to uncover some unexpected disorder-controlled phenomena.

The key to our solution lies in first writing the backward equation for the moment generating function of the first-passage probability, and then reformulating the resulting recursion as a linear algebra problem. This approach leads to analytic expressions for the generating function in terms of the elements of the inverse of a tridiagonal matrix~\cite{usmani1994inversion}. Related methods have previously provided a semi-analytic solution to the probability distribution from one-step master equation in time~\cite{smith2015general}, and to solve the three-term recurrence for the moment generating function~\cite{holehouse2023recurrence}. Other pertinent studies derive semi-analytic solutions for the first-passage time distribution, again for arbitrary hopping rates~\cite{barrio2013reduction,leier2014exact,ashcroft2015mean}. These studies start with the Laplace transform of the formal solution to the master equation, and obtain their result in terms of undetermined eigenvalues of the master operator.  

%Perhaps the closest study to ours uses a generating function approach to treat first-passage on the disordered interval~\cite{noskowicz1990first} to compute the mean first-passage time and its variance over different realizations of disorder. 

However, these past investigations all focused on the forward master equation, whereas we use the backward equation due to its utility in the study of first-passage problems~\cite{redner2001guide,iyer2016first}. The benefits of our approach are its versatility in elucidating first-passage properties for absorbing boundaries, reflecting boundaries, and conditional waiting times, and its conciseness through the use of linear algebraic results that simplify the generating function. Additionally, our approach does not rely on finding the eigenvalues of the master operator numerically, leading to analytic, as opposed to semi-analytic results.

\smallskip\emph{Formalism.} The backward equation is the adjoint of the commonly used master equation for the time evolution of the probability distribution.  The backward equation is especially useful in first-passage problems for which the final state is prescribed and the initial state becomes the fundamental dependent variable. Let $P_{i,j}(t)$ be the probability for a random walk to reach a final state $j$ for the \emph{first} time at time $t$ when starting from state $i$ at $t=0$.  Here, $i$ denotes a one-dimensional coordinate. The backward equation for this first-passage probability is~\cite{van1992stochastic,redner2001guide,ashcroft2016metastable},
\begin{align}
\label{eq:bwMEraw}
    P_{i,j}(t+\Delta t)&= b_i\Delta t P_{i+1,j}(t)+ d_i\Delta t P_{i-1,j}(t) \nonumber\\
    &\hspace{1.3cm} +[1-(b_i+d_i)\Delta t]P_{i,j}(t),
\end{align}
where $b_i$ and $d_i$ are arbitrary rates of hopping to the right and left from site $i$, respectively. This equation states that the probability to first arrive at $j$ starting from $i$ at time $t+\Delta t$ equals the sum of the hopping probabilities in a time $\Delta t$ from $i$ to either $i-1$, $i$, or $i+1$ times the first-passage probabilities to $j$ at time $t$ from $i-1$, $i$, or $i+1$. We set $\Delta t=1$ so that the hopping probabilities in a time $\Delta t$ are $b_i\Delta t$ and $d_i\Delta t$, and $t$ now counts the number of hopping events. For $\Delta t=1$, the hopping rates must satisfy the constraints $0<b_i,d_i<1$ and $b_i+d_i<1$.

\begin{figure}[ht]
    \includegraphics[width=.375\textwidth]{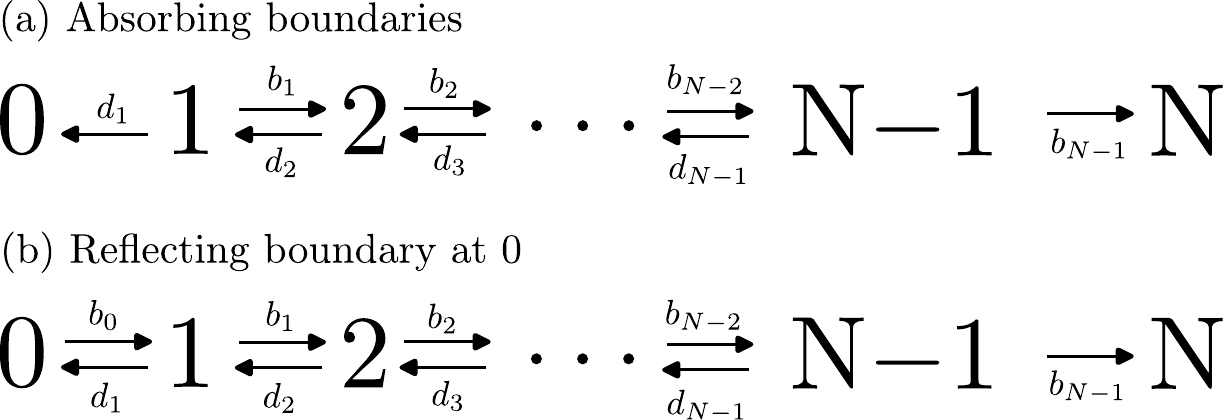}
    \caption{Interval with (a) two absorbing boundaries, and (b) one reflecting and one absorbing boundary.}
    \label{fig:fig0}
\end{figure}

We now calculate the unconditional first-passage probability for the interval with two absorbing boundaries (Fig.~\ref{fig:fig0}(a)). Define $f_i(t)=P_{0,i}+P_{N,i}$ as the probability density function to first reach either $0$ or $N$ at time $t$ when starting from site $i$.  In close analogy with Eq.~\eqref{eq:bwMEraw}, its governing equation is
\begin{align}\label{eq:feq}
    f_i(t\!+\!1)= b_i f_{i+1}(t) + d_i f_{i-1}(t) +[1\!-\!(b_i\!+\!d_i)]f_i(t).
\end{align}
We solve this equation by the generating function technique. Multiplying \eqref{eq:feq} by $z^{t}$, summing over $t$, and introducing the generating function $F_i(z) = \sum_{t=0}^\infty z^{t}f_i(t)$, gives the following three-term recurrence
\begin{align}
\label{eq:RRforMGFuncond}
    \begin{split}
    &b_i F_{i+1}(z)+d_i F_{i-1}(z)+\beta_i(z) F_i(z)=0,
    \end{split}
\end{align}    
for $1\leq i\leq N-1$, where for notational simplicity we introduce $\beta_i(z)\equiv 1-z^{-1}-b_i-d_i$.
This recursion obeys the boundary conditions $F_0(z) = F_N(z) = 1$, since $f_0(t)=f_N(t) = \delta_{t,0}$. 

%From this, we can readily obtain the factorial moments by
%\begin{align}\label{eq:moms-calc}
%    \mathbb{E}_i[t(t-1)\ldots(t-m+1)] = \partial_z^m F_i(z)|_{z\to 1^-}\,,
%\end{align}
%while the FPT distribution itself is given by
%\begin{align}\label{eq:prob-calc}
%    f_i(t) = \frac{1}{t!}\partial^t_{z}F_i(z)|_{z\to 0^+}\,.
%\end{align}

To solve \eqref{eq:RRforMGFuncond}, it is helpful to write it as a matrix equation~\cite{holehouse2023recurrence}. We first define the column vector $\mathbf{F}(z)=(F_1(z),F_2(z),\dots,F_{N-1}(z))$ and using $F_0(z) = F_N(z) = 1$, this recursion becomes
\begin{align}
\label{eq:MGFmatEq1}
    \mathbb{A}(z)\cdot \mathbf{F}(z) = -\mathbf{w},
\end{align}
where  $\mathbf{w} = (d_1,0,0,\dots,0,b_{N-1})$ and $\mathbb{A}(z)$ is the tridiagonal matrix of dimension $(N-1)\times(N-1)$
\begin{align*}
\label{eq:Adef}
    \mathbb{A}(z)=
    \begin{pmatrix}
        \beta_1(z) & b_1 & 0 & \dots & 0 & 0 \\
        d_2 & \beta_2(z) & 0 & \dots & 0 & 0 \\
        \vdots &  &  &  & \ddots &  &  \\
        0 & 0 & 0 & \dots & d_{N-1} & \beta_{N-1}(z) \\
    \end{pmatrix}.
\end{align*}
The formal solution to Eq.~\eqref{eq:MGFmatEq1} is,
\begin{align}
    \mathbf{F}(z) = -\mathbb{A}(z)^{-1}\cdot \mathbf{w}.
\end{align}
In what follows we write the $ij^{\rm th}$ elements of $\mathbb{A}(z)^{-1}$ as $\alpha_{i,j}(z)$. Performing the matrix multiplication gives 
\begin{align}
    F_i(z) = -[d_1\alpha_{i,1}(z)+b_{\bar{N}}\alpha_{i,\bar{N}}(z)]\,,
\end{align}
with $\bar{N}\equiv N-1$. Our task now is to find closed-from expressions for the elements of the inverse $\alpha_{i,j}(z)$. To this end we exploit the tridiagonal form of $\mathbb{A}(z)$, as was done in~\cite{smith2015general} for the forward master equation, to give the elements $\alpha_{i,j}(z)$ in terms of computationally simple products of polynomials via Cramer's rule~\cite{usmani1994inversion}. For later use, we define the following products
\begin{align}\label{eq:BDdef}
    B_i \equiv \prod_{k=i}^{\bar{N}} b_k,\qquad D_i \equiv \prod_{k=1}^i d_k.
\end{align}
To find $F_i(z)$, we only require the elements $\alpha_{i,1}(z)$ and $\alpha_{i,\bar{N}}(z)$, which are given by
\begin{align}\label{eq:alphas}
    \alpha_{i,1} = 
    (-1)^{i+1}\frac{D_i p_{i}(z)}{d_1 p_0(z)},\quad
    \alpha_{i,\bar{N}} = (-1)^{i+\bar{N}}\frac{B_i q_{i-1}(z)}{b_{\bar{N}} p_0(z)},
\end{align}
where we recursively define the polynomials
\begin{subequations}
\label{eq:pq-def}
\begin{align}
%\label{pn-def}
\begin{split}
    p_{N}(z)&=1,\\
    p_{\bar{N}}(z)&=\beta_{\bar{N}} (z),\\  
    p_{i}(z) &= \beta_{i+1}(z)p_{i+1}(z)-b_{i+1}d_{i+2}p_{i+2}(z),
\end{split}    
\end{align}
for $0 \leq i\leq \bar{N}-2$, and
\begin{align}
%\label{qn-def}
\begin{split}
%\label{eq:lastOrtho}
    q_0(z)&=1,\\
    q_1(z) &= \beta_1(z),\\
    q_i(z) &= \beta_i(z)q_{i-1}(z)-d_i b_{i-1}q_{i-2}(z)\,,
\end{split}
\end{align}
\end{subequations}
for $2\leq i\leq \bar{N}$. Although this calculation seems complicated, it is much faster than matrix inversion via Cramer's rule, since it avoids calculating the multiple zeros encountered in the evaluation of the minors of $\mathbb{A}(z)$~\cite{higham2002accuracy}.  More details are given in Appendix~\ref{app:app1}.
% To help the intuition of the reader, we have shown the calculation of some of the matrix elements in Appendix \ref{sec:AppMatInv}, where we also motivate the definition of the orthogonal polynomials above. The determinant of the matrix is equivalent to $\mathrm{det}(\mathbb{A}(z))=p_0(z)=q_{\bar{N}}(z)$. 

We can now write the generating functions for the first-passage probability for each starting position $i$ as
\begin{align}
\label{eq:uncond-GF}
\begin{split}
    F_1(z) &= \frac{1}{p_0(z)}\left[(-1)^{\bar{N}}B_1 -d_1p_{1}(z)\right],\\
       F_{\bar{N}}(z) &= \frac{1}{p_0(z)}\left[ (-1)^{\bar{N}}D_{\bar{N}} - b_{\bar{N}} q_{\bar{N}-1}(z) \right],\\
     F_i(z) &= \frac{(-1)^{i}}{p_0(z)}\left[ p_{i}(z)D_i  + (-1)^{\bar{N}-1} q_{i-1}(z)B_i  \right]\,,
\end{split}
\end{align}
where the last equality holds for $2\leq i\leq\bar{N}-1$. Note that $p_{\bar{N}-i}$ and $q_i$ are polynomials of order $i$ in $z^{-1}$, and each $F_i(z)$ is a rational function that is made up of polynomials in $z$. In practice, for an interval of length $N$ and given set of $\{b_i,d_i\}$, we find the polynomials $p_i$ and $q_i$ via Eq.~\eqref{eq:pq-def}, after which we can use Eqs.~\ref{eq:uncond-GF} to compute the generating function and its series expansion. 

Our formalism can be readily extended to the \emph{conditional} first-passage probability, namely, the probability to first reach a specified boundary without ever touching the other boundary (Appendix~\ref{sec:condWT}). We can also treat the case of a reflecting boundary. Here, the dynamics at the reflecting end of the interval must be treated as a special case in which some of the elements of $\mathbb{A}(z)$ are altered, although the functional forms in Eqs.~\ref{eq:uncond-GF} remain the same.  The details of this calculation will be given elsewhere.

\smallskip\textit{Application to the disordered interval:} A salient feature of first-passage in the interval is the huge disparity in the mean first-passage time (FPT) between different realizations of the hopping rates $\{b_j,d_j\}$.  We use Eq.~\eqref{eq:uncond-GF} to give the exact FPT that represents an average over all random-walk trajectories. We choose each $b_j$ from a uniform distribution on  $[0,2/3]$, so that each $d_j=2/3-b_j$.  Since $b_j+d_j=2/3$, there is a 1/3 probability for the walk to remain at the same site in a single event.  This choice eliminates the even-odd oscillations that arise for the nearest-neighbor random walk with $b_j+d_j=1$. 

\begin{figure}[ht]
\vspace{-4mm}
\centerline{\hspace{8mm}\includegraphics[width=.42\textwidth]{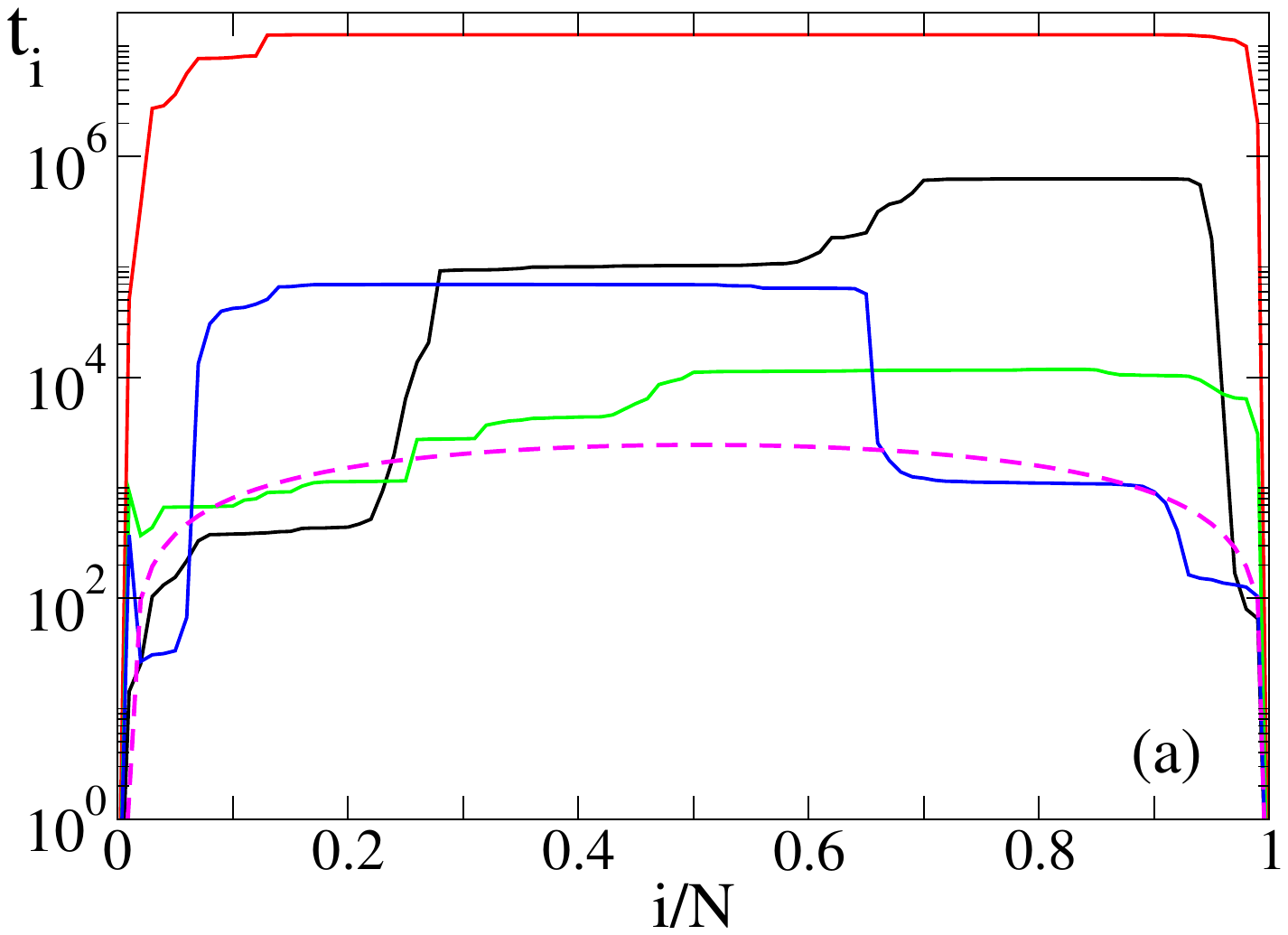}}
\centerline{\includegraphics[width=.38\textwidth]{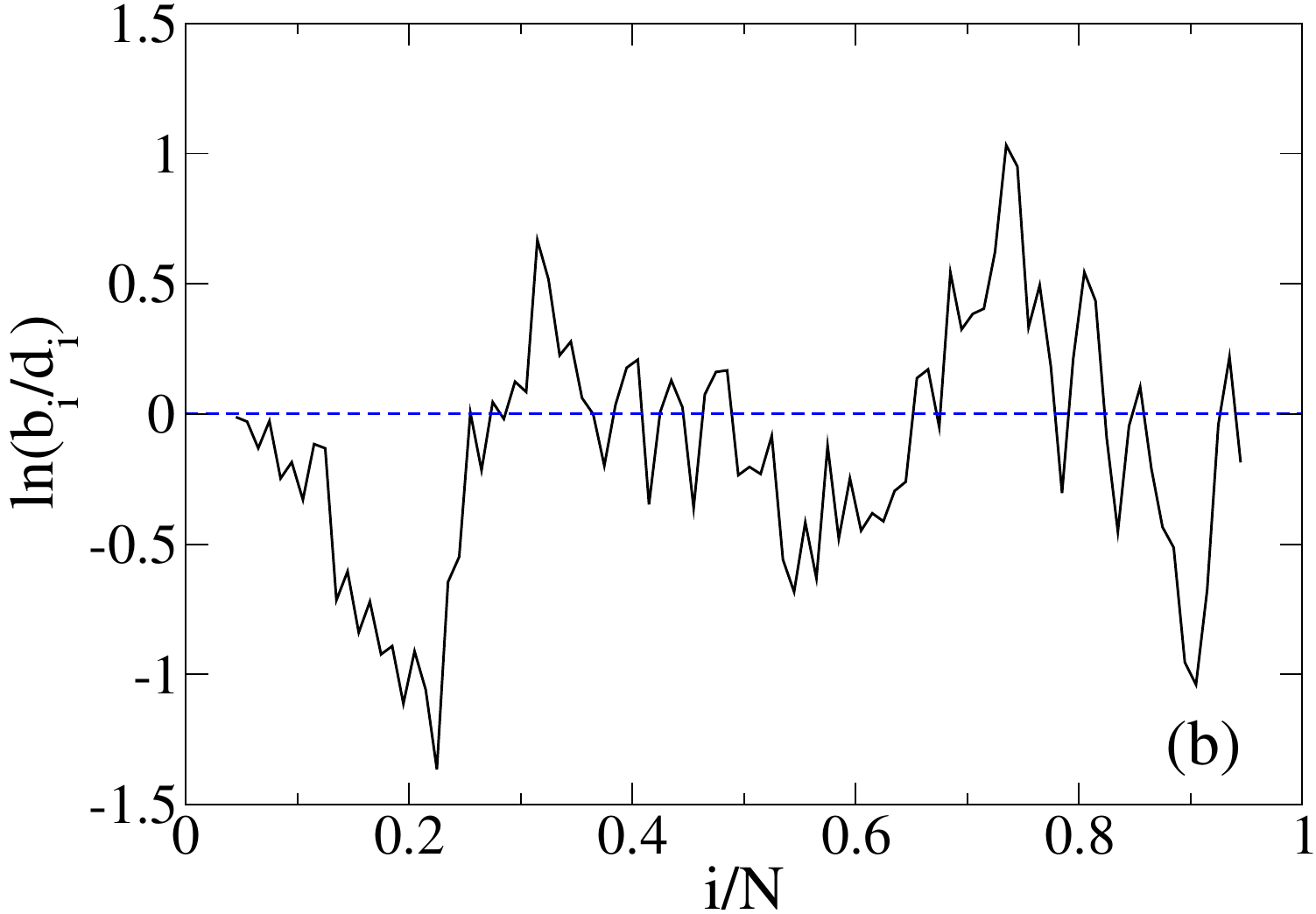}}
    \caption{(a) Mean first-passage times on a disordered interval of length $N=100$ for four realizations of the hopping rates as a function of starting position $i$ (solid).  The dashed curve shows the first-passage time for the homogeneous interval with $b_i=d_i=1/3$ for all $i$. (b) The smoothed (10-point range) local bias $\ln(b_i/d_i)$ for the realization in black in (a).}
    \label{fig:fig1}
\end{figure}

Figure~\ref{fig:fig1} illustrates this heterogeneity in the FPT for several realizations of the $\{b_j,d_j\}$; even larger realization-specific variation occurs for higher moments of the FPT (not shown). Moreover, the dependence of the FPT on starting position has no resemblance to the parabolic profile that arises in the absence of disorder. Instead, the FPT is nearly independent of starting location in certain subintervals and changes rapidly within intervening boundary layers.  This behavior stems from the existence of local potential wells that are induced by the disordered hopping rates. For the realization shown in black in the figure, there is a negative bias over the first quarter of the interval so that the FPT for starting points in this range is small.  The bias suddenly becomes positive for $i/N\agt 0.3$ which causes the rapid increase in the FPT.  The local bias again changes sign as $i/N$ increases through roughly 0.65.  The effective potential well for $i/N$ between 0.3 and 0.65 keeps the FPT roughly constant in this range. The change of sign in the bias around $i/N\approx 0.7$ is the source of another sudden increase in the FPT at this point. Generally, wherever the local bias leads to an effective potential well, the FPT is nearly independent of starting location within this well and then suddenly jumps when the local bias changes sign.

It is also revealing to determine how the FPT converges to its true value upon averaging over progressively larger numbers of realizations of the hopping rates.  For this calculation, we use the dichotomous distribution in which each $b_j$ takes the values 0.3 or 0.6 equiprobably, while $d_j=0.9-b_j$.  With this choice, there is a countably finite number of hopping rate realizations so that we can average over all random walk trajectories \emph{and} over all realizations of the hopping rate disorder for a given (albeit short) interval length.

\begin{figure}[ht]
    \includegraphics[width=.45\textwidth]{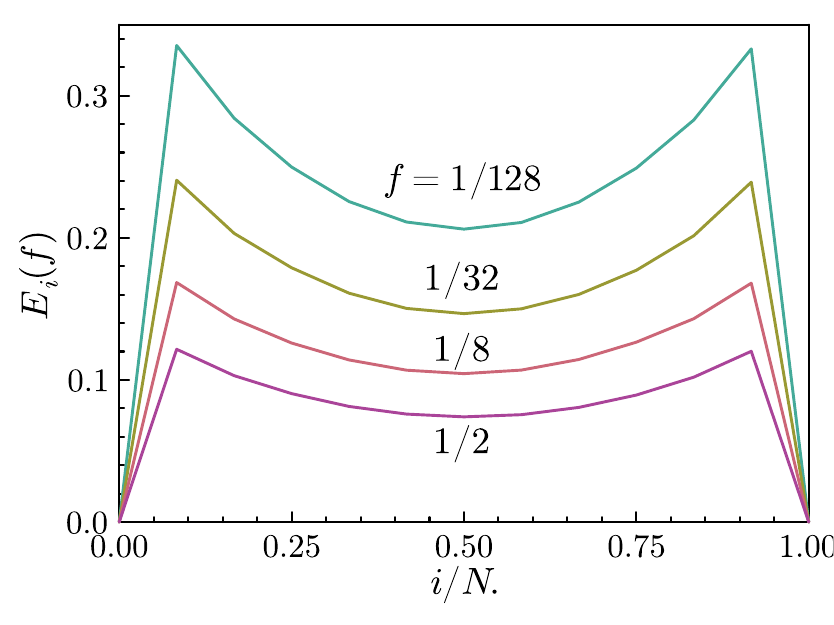}
    \caption{The average deviation from Eq.~\eqref{eq:Ef} when a finite fraction $f$ of all hopping rate realizations are sampled. Interval length $N=12$.}
    \label{fig:fig2}
\end{figure}

For a given realization of the hopping rates indexed by $\alpha$, we define the mean FPT starting from site $i$ as $\langle t_i^{(\alpha)}\rangle$. The true average FPT, averaged over all $\mathcal{M}=2^{N-1}$ realizations of the hopping rates is then
\begin{align*}
\overline{\langle t_i\rangle} \equiv \frac{1}{\mathcal{M}}\sum_{\alpha=1}^\mathcal{M} \langle t_i^{(\alpha)}\rangle\,.
\end{align*}
We now ask: how close is the average FPT over a subset of realizations of the hopping rates to the average over \emph{all} realizations of the hopping rates? To address this question, we define the partial average in which we select a random fraction $f=M/\mathcal{M}$ of all hopping rate realizations and compute the exact FPT for this subset, again for random walks that start at site $i$:
\begin{align*}
\overline{\langle t_i\rangle}_f \equiv \frac{1}{M}\sum_{\alpha=1}^M \langle t_i^{(\alpha)}\rangle\,.
\end{align*}
An appropriate deviation measure is the relative difference between the true average and the partial average as a function of $f$, which we define as $E_i(f)$:
\begin{align}
\label{eq:Ef}
    E_i(f) \equiv \frac{ \overline{\langle t_i\rangle} -\overline{\langle t_i\rangle}_f }{ \overline{\langle t_i\rangle}}\,.
\end{align}

\begin{figure}[ht]
    \includegraphics[width=.4\textwidth]{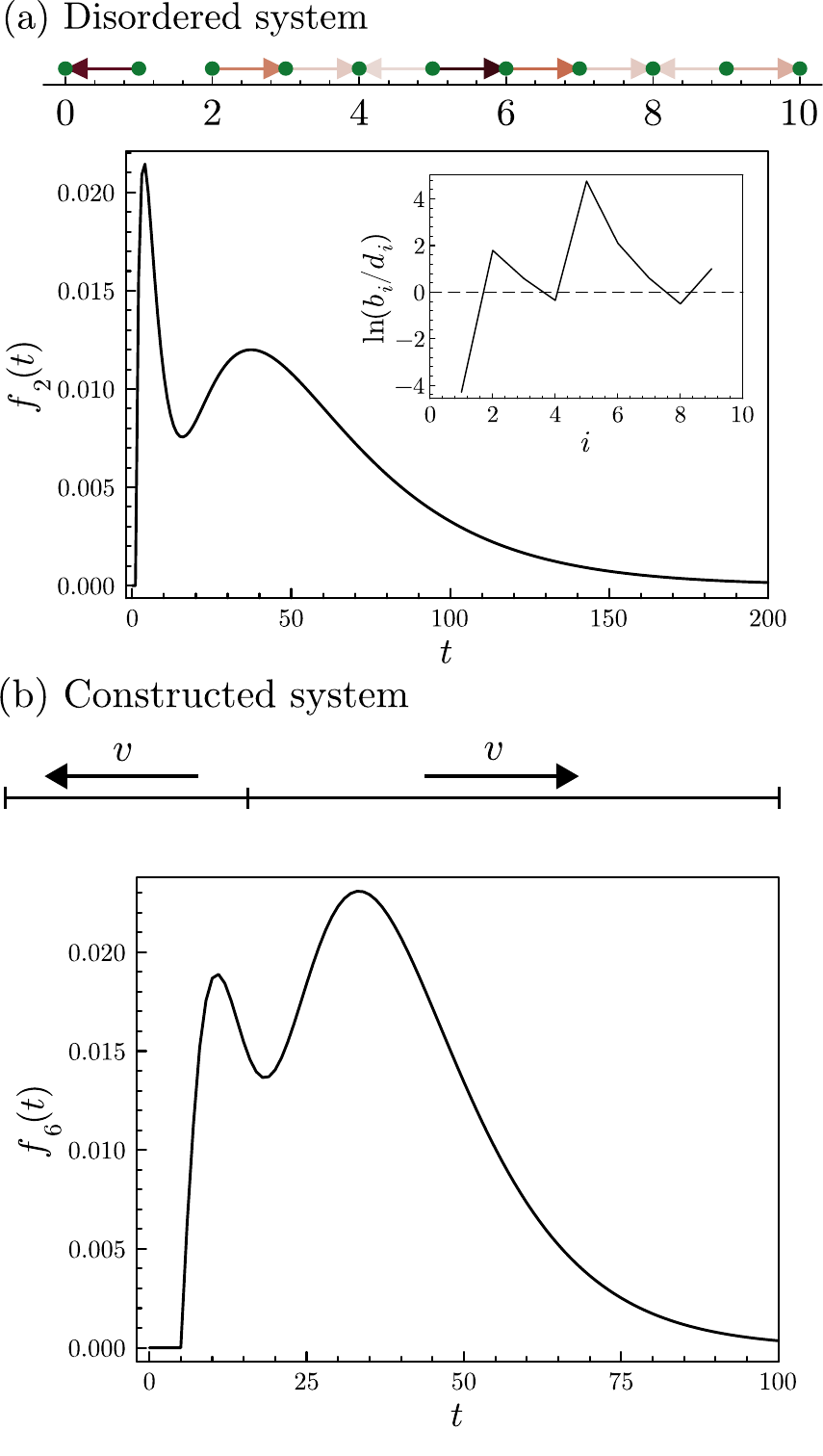}
    \caption{(a) The first-passage probability $f_{2}(t)$ on the interval $[0,10]$. The schematic shows the local bias $\ln(b_i/d_i)$ at each site (darker color = stronger bias). The inset shows the explicit values of $\ln(b_i/d_i)$. (b) The first-passage probability $f_{6}(t)$ for an interval of length $N=20$ in which the synthetically generated bias changes sign at $i=5$ with $|v|=|b_i-d_i|=0.3$. }
    \label{fig:fig3}
\end{figure}

In Fig.~\ref{fig:fig2}(a) shows this deviation $E_i(f)$ versus $i$ for various ensemble fractions $f$.  We see that with 1\% of all realizations there is a deviation of roughly 25\%.  Since a real simulation can only sample an infinitesimal fraction of all realizations, this plot indicates that any reasonable simulation will provide a poor approximation to the true average FPT. In particular, having half of all possible realizations still leads to a deviation of around 10\%. We conclude that studies of first-passage in disordered systems are effectively pointless from a simulation perspective. Again, the deviations found for the higher-order moments are more substantial than that of the mean behavior (not shown).

Another unexpected feature from our exact approach is that \emph{bimodal} first-passage distributions arise for certain realizations of the hopping rates. Such effects can be difficult to observe from a simulation-based approach as multiple modes can seem apparent due to finite-sample noise. We show one such example in Fig.~\ref{fig:fig3}(a) with $b_i$ chosen uniformly in the range $(0,1/3)$ and $d_i=1/3-b_i$, with the walk starting at $i=2$ on an interval of length 10. Also shown in this figure are the local biases at each site $i$, $\ln(b_i/d_i)$.  The essential feature of this bias profile is that it is negative at $i=1$, almost neutral at $i=2$, and generally positive for $i>2$.  Thus a particle starting at $i=2$ exits via the left edge with appreciable probability and does so quickly.  However, if the particle initially hops to the right, it then experiences a rightward bias, which leads to the second peak in the first-passage probability. To verify this crude picture, we also construct a synthetic interval of length 20 in which the segment $[0,5]$ has a bias to the left and the segment $[5,20]$ has a bias to the right both of magnitude $|v| = 0.3$ (Fig.~\ref{fig:fig3}(b)).  When the random walk starts at $i\leq 5$, it is likely to remain in the region $i\leq 5$ and exit the interval on the left side, corresponding to the early-time peak in Fig.~\ref{fig:fig3}(b). However, once the random walker traverses to the region $i>5$, it is likely to remain on this side of the interval, ultimately leading to the second, longer-time peak in Fig.~\ref{fig:fig3}(b). 

\textit{Summary:} We derived the first-passage probability and all its moments on disordered intervals with arbitrary nearest-neighbor hopping rates.  Our approach gives a simple way to understand various anomalies, such as the step-like features of the first-passage time for individual hopping-rate realizations.  We also quantified how the first-passage time averaged over finite fraction of all realizations of the hopping rates approaches the true average when averaging over 
all realizations of the hopping rates.  In any realistic simulation of a disordered chain, it is practical to sample only a tiny fraction of all realizations of the disorder, and our results show that this limitation severely compromises the accuracy of such simulations. We additionally found the unexpected feature of bimodal first-passage distributions.  These arise in hopping-rate realizations that have a general outward bias from the interval. For such a system, a random walk exits via one side of the interval quickly and more slowly from the other side, but with comparable probabilities for each mode of exit. It would interesting to see whether this bimodality has potentially useful applications in describing unexplained occurrences of multimodality in other stochastic processes. In particular, the relation between transient bimodality in the solution to the forward equation, where bimodality can arise and dissipate on the way to the steady state, and bimodalities in the first-passage time distribution have not been widely explored \cite{broggi1985transient,lange1985study,jia2020dynamical,holehouse2020stochastic}. 
%Finally, extending these results to dimensions higher than 1 would be of great utility, since the issues of glassiness and long metastable behaviors only become more complex in multiple dimensions.

\emph{Acknowledgements:} This publication is based upon work that is supported by the National Science Foundation under Grant No.~DMR-1910736.

% \section*{Acknowledgements}

%\sid{I think the rest of this paragraph is both too vague and too weak to stand as is.} Future avenues include the possibility of using the analytics herein to extract asymptotic results that scale better when finding the series coefficients of the generating function. We note that the issue of having a generating function whose series coefficients are difficult to determine is not confined to the present study, but extends to cases where generating functions include products of hypergeometric, Heun \cite{ronveaux1995heun} or other complicated special functions with other lower-order functions \cite{NIST:DLMF} (e.g., \cite{grima2012steady} and \cite{iyer2009stochasticity}). This issue means that increased computational power is required for series expansions to higher-orders and higher values of the system size $N$, and means that conclusions of our findings in the thermodynamic limit cannot be accurately made. 

%how one can artificially create first-passage times with multiple modes based on the findings observed for the disordered interval, and 

% subsection practical_evaluation_of_ (end)
\bibliographystyle{naturemag.bst}
\bibliography{main}

\pagebreak
\clearpage
\widetext
\begin{center}
\textbf{\large Supplementary Information}
\end{center}
%%%%%%%%%% Merge with supplemental materials %%%%%%%%%%
%%%%%%%%%% Prefix a "S" to all equations, figures, tables and reset the counter %%%%%%%%%%
\setcounter{equation}{0}
\setcounter{figure}{0}
\setcounter{table}{0}
\setcounter{footnote}{0}
\setcounter{page}{1}
\makeatletter
\renewcommand{\theequation}{S\arabic{equation}}
\input{appendix}

\end{document}

%% file: appendix.tex
\section{Inverse elements of a tridiagonal matrix}\label{app:app1}
In this appendix we succinctly reproduce  the inverse elements of a general tridiagonal matrix first given in \cite{usmani1994inversion}. That is, we will derive the elements of the inverse of the following matrix,
\begin{align}\label{eq:Adef}
    \mathbb{B}=
    \begin{pmatrix}
        b_1 & c_1 & 0 & 0 & \dots & 0 & 0 \\
        a_2 & b_2 & c_2 & 0 & \dots & 0 & 0 \\
        0 & a_3 & b_3 & c_3 & \dots & 0 & 0 \\\vdots &  &  &  & \ddots &  &  \\
        0 & 0 & 0 & 0 & \dots & a_{N} & b_N \\
    \end{pmatrix},
\end{align}
First, we know that the inverse of a general $N\times N$ matrix is given by Cramer's rule as \cite{higham2002accuracy}, 
\begin{align}\label{eq:alpha}
    \alpha_{ij}\equiv[\mathbb{B}^{-1}]_{i,j} = \frac{(-1)^{i+j}M_{ji}}{\textrm{det}(\mathbb{B})},
\end{align}where $M_{ji}$ is the minor, i.e., the determinant of the matrix that results from the removal of row $j$ and column $i$. Our aim is to exploit the triangular nature of the matrix to give analytic expressions for the inverse that can be computed much faster than general matrix inversion (which scales as $N^3$ for Gauss--Jordan elimination). For brevity we introduce the following determinants,
\begin{align}
    U_i \equiv \begin{vmatrix}
                b_1 & c_1 & 0 & 0 & \dots & 0 & 0 \\
                a_2 & b_2 & c_2 & 0 & \dots & 0 & 0 \\
                0 & a_3 & b_3 & c_3 & \dots & 0 & 0 \\\vdots &  &  &  & \ddots &  &  \\
                0 & 0 & 0 & 0 & \dots & a_i & b_i \\
          \end{vmatrix},\qquad\qquad
    L_i \equiv \begin{vmatrix}
                b_i & c_i & 0 & 0 & \dots & 0 & 0 \\
                a_{i+1} & b_{i+1} & c_{i+1} & 0 & \dots & 0 & 0 \\
                0 & a_{i+2} & b_{i+2} & c_{i+2} & \dots & 0 & 0 \\\vdots &  &  &  & \ddots &  &  \\
                0 & 0 & 0 & 0 & \dots & a_N & b_N \\
          \end{vmatrix},
\end{align}
with $U_0=1$, $U_1 = b_1$ and $L_{N+1} = 1$, $L_N = b_N$. The key step in calculating the $\alpha_{ij}$ is in finding the matrix minors. Let's calculate the minors for a few examples,
\begin{align}\nonumber
    M_{11} &= L_2,\\\nonumber
    M_{22} &= U_1 L_3,\\\nonumber
    M_{33} &= U_2 L_4,\\\nonumber
    M_{12} &= a_2 L_3,\\\nonumber
    M_{13} &= a_2 a_3 L_4,\\\nonumber
    M_{14} &= a_2 a_3 a_4 L_5,\\\nonumber
    M_{N,N} &= U_{N-1},\\\nonumber
    M_{N,N-1} &= c_{N-1} U_{N-2},\\\nonumber
    M_{N,N-2} &= c_{N-1}c_{N-2} U_{N-3}.
\end{align}
Here we have made judicious use of Schur's formula \cite{zhang2006schur} for the determinants of block matrices, which states that for the block matrix
\begin{align}
    \mathbb{E} = \begin{pmatrix}
        \mathbf{A} & \mathbf{B} \\
        \mathbf{C} & \mathbf{D}
    \end{pmatrix},
\end{align}
with invertible $\mathbf{A}$ and $\mathbf{D}$, we have
\begin{align}
    \mathrm{det}(\mathbb{E}) = \mathrm{det}(\mathbf{A})\cdot\mathrm{det}(\mathbf{D} - \mathbf{C}\cdot \mathbf{A}^{-1}\cdot \mathbf{B}).
\end{align}
When either $\mathbf{B}$ or $\mathbf{C}$ consist entirely of zeros, the above formula reduces to,
\begin{align}
    \mathrm{det}(\mathbb{E}) = \mathrm{det}(\mathbf{A})\cdot\mathrm{det}(\mathbf{D}).
\end{align}
We observe that the minors follow the pattern,
\begin{align}
    M_{ji} = 
    \begin{cases}
        \left(\prod_{k=i}^{j-1}c_k\right)U_{i-1}L_{j+1}& j>i,\\[2mm]
        U_{j-1}L_{j+1}& j=i,\\[2mm]
        \left(\prod_{k=j+1}^{i}a_k\right)U_{j-1}L_{j+1}& j<i,\\
    \end{cases} 
\end{align}
which is proved by induction in \cite{usmani1994inversion2}. Substituting $M_{ji}$ into Eq.~\eqref{eq:alpha} then gives our analytic expression for the $\alpha_{ij}$. Note that $\det(\mathbb{B}) = L_1 = U_N$. 

The introduction of orthogonal polynomials comes from the recursive relationship defining the determinants of the sequences $U_0,U_1,\ldots,U_N$ and $L_1,L_2,\ldots,L_{N+1}$. This recurrence is trivially given by Leibniz's rule as,
\begin{align}
\begin{split}
    L_{N+1} &= 1,\; L_N = b_N,\\
    L_i &= b_i L_{i+1} - c_i a_{i+1} L_{i+2},\\
    U_0&=1,\; U_1 = b_1,\\
    U_i &= b_i U_{i-1} - c_{i-1}a_{i}U_{i-2}.
\end{split}
\end{align}
Upon re-labeling $b_i\Rightarrow \beta_i(z)$, $a_i\Rightarrow d_i$, $c_i \Rightarrow b_i$, $N\Rightarrow \bar{N}$, $U_i \Rightarrow q_i(z)$ and $L_i \Rightarrow p_{i-1}(z)$ one then recovers the recursive polynomials defined in Eq.~\eqref{eq:pq-def} in the main text. Finally, since we only need the terms $\alpha_{i,1}(z)$ and $\alpha_{i,\bar{N}}(z)$, we can use the properties of $U_0=1$ and $L_{N+1}=1$ to arrive at Eqs.~\eqref{eq:alphas} in the main text.

\section{Conditional first-passage probability}\label{sec:condWT}
We are interested in the \emph{conditional} FPT to reach either $0$ or $N$ without ever reaching the opposite boundary starting from some initial site $i$. We focus on the conditional FPT to reach $N$, and later return to the case of reaching the origin.  The probability of reaching $N$ at time $t$ starting from $i$ at $t=0$ is given by $f_i(t) = P_{i,N}(t)$. Note that $f_i(t)$ has the boundary conditions $f_N(t)=\delta_{t,0}$ and $f_0(t)=0$. The recurrence relation satisfied by $f_i(t)$ is the same as Eq.~\eqref{eq:feq}, but with different boundary conditions. Note that $f_i(t)$ does not strictly correspond to a normalized probability distribution over $t$ since there is a finite probability of being absorbed at the origin and therefore never reaching $N$. Therefore, we further define,
\begin{align}
    \widetilde{f}_i(t) = \frac{f_i(t)}{\sum_{t=0}^\infty f_i(t)}\equiv \frac{f_i(t)}{\phi_i},
\end{align}
where $\widetilde{f}_i(t)$ is the properly normalized probability density function and we define $\phi_i$ as the probability for the walk to ultimately reach the boundary at $N$. This eventual hitting probability is given by
\begin{align}
    \phi_i = \frac{1+\sum_{k=1}^{i-1}\prod_{i=1}^k \frac{d_j}{b_j}}{1+\sum_{k=1}^{N-1}\prod_{i=1}^k \frac{d_j}{b_j}}.
\end{align}
We now define the generating function of $\widetilde{f}_i(t)$ as,
\begin{align}
    \widetilde{F}_i(z) = \sum_{t=0}^\infty z^{t}\widetilde{f}_i(t),
\end{align}
and taking Eq.~\eqref{eq:feq} multiplying by $z^t$ and summing over all $t$ we get the three-term recurrence relation that $\widetilde{F}_i(z)$ satisfies,
\begin{align}\label{eq:RRforMGFcond}
    b_i \phi_{i+1}\widetilde{F}_{i+1}(z)+d_i \phi_{i-1}\widetilde{F}_{i-1}(z)+\beta_i(z)\phi_i \widetilde{F}_i(z)=0,\qquad i\in{1,2,\dots,N-1},
\end{align}
with the boundary conditions now $\widetilde{F}_0(z)=0$ and $\widetilde{F}_N(z)=1$ and we have used $\beta_i(z)$ as defined in the main text. For brevity, we henceforth denote $C_i(z) = \phi_i\widetilde{F}_i(z)$. As in the unconditional case, we rewrite Eq.~\eqref{eq:RRforMGFcond} as a matrix equation,
\begin{align}
    \mathbb{A}(z)\cdot \mathbf{C}(z) = -\mathbf{m},
\end{align}
with $\mathbf{C}(z) = (C_1(z),C_2(z),\dots,C_{\tau}(z))$, $\mathbf{m}=(0,0,\dots,0,b_{\tau})$ and $\mathbb{A}(z)$ the $(N-1)\times(N-1)$ matrix defined by Eq.~\eqref{eq:Adef}. Since we already have explicit expressions for the elements of the inverse of $\mathbb{A}$, we can readily obtain $\widetilde{F}_i(z) = -b_{\bar{N}} \alpha_{i,\tau}/\phi_i$:
\begin{align}\label{eq:cond-GF-1}
\begin{split}
    \widetilde{F}_i(z) &= \frac{-(-1)^{i+\tau}q_{i-1}(z)B_i}{p_0(z) \phi_i},\; \qquad i\in\{ 1,2,\dots,\bar{N}-1 \},\\[2mm]
    \widetilde{F}_{\bar{N}}(z) &= \frac{-b_{\bar{N}} q_{-1}(z) }{p_0(z)\phi_{\bar{N}}}.
\end{split}
\end{align}
By symmetry considerations one can then find the conditional generating functions of the first-passage time to reach $n=0$ as,
\begin{align}\label{eq:cond-GF-2}
\begin{split}
    \widetilde{F}_1(z) &= \frac{-d_1p_{1}(z)}{p_0(z)(1-\phi_1)},\\[2mm]
    \widetilde{F}_i(z) &= \frac{(-1)^{i}p_{i}(z)D_i}{p_0(z)(1-\phi_i)},\; \qquad i\in\{ 2,3,\dots,\bar{N} \}.
\end{split}
\end{align}
In the application of the disordered interval in the main text one can use the conditional first-passage time to observe the origin of bimodalities observed for some realizations of disorder. This is possible since the origin of the bimodality seen in Figure \ref{fig:fig3} comes from the competition between leaving the opposing ends of the system.